\documentclass[pre,aps,twocolumn,showpacs,floatfix]{revtex4}
\usepackage{graphicx}
\usepackage{amsmath, amsthm, amssymb}
\begin{document}

\title{Pressure exerted by a grafted polymer: Bethe lattice solution}
\author{Rafael Mynssem Brum}
\author{J\"{u}rgen F. Stilck}
\email{jstilck@if.uff.br}
\affiliation{Instituto de F\'{i}sica and National Institute of Science
and Technology for Complex Systems, Universidade Federal Fluminense, Av.
Litor\^anea s/n, 24210-346 Niter\'oi, RJ, Brazil}

\date{\today}

\begin{abstract}
We solve the problem of a chain, modeled as a self-avoiding walk,  grafted o the wall limiting a semi-infinite Bethe lattice of arbitrary coordination number $q$. In particular, we determine the pressure exerted by the polymer on the wall, as a function of the distance to the grafting point. The pressure, in general, decays exponentially with the distance, at variance with what is found for SAWs and directed walk on regular lattices and gaussian walks. The adsorptions transition, which is discontinuous, and its influence on the pressure are also studied.
\end{abstract}

\pacs{05.50.+q,36.20.Ey,05.50.Fh}

\maketitle

\section{\label{intro}Introduction}
The problem of the non-homogeneous pressure applied by a polymeric chain to the wall to which it is grafted, besides having some interest as a basic problem in polymer physics, is also related to the deformation of a biological membrane to which a protein molecule is attached \cite{bmj00}. Here we consider the version of the model where the wall is rigid, and therefore does not deform. The continuous version of this problem, where a $d$-dimensional semi-infinite space is limited by a $(d-1)$-dimensional wall, was solved for the case where the polymer is modeled as a gaussian walk, by Bickel, Marques and Jeppesen \cite{bmj00}. The dimensionless entropic pressure applied by the gaussian walk to a point on the wall at a distance $r$ of the grafting point is:
\begin{equation}
p_G(r)=\frac{P_G(r)a^d}{k_BT}=\frac{\Gamma(d/2)}{\pi^{d/2}}
\frac{1}{(r^2+1)^{d/2}},
\label{gauss}  
\end{equation}
where the distance $r$ is measured in units of the distance between consecutive monomers of the chain $a$. The lattice version of this problem was studied using exact enumerations of SAWs on a semi-infinite square lattice \cite{j13}.  

In the canonical formalism for a simple fluid, the pressure may be obtained from the fundamental equation for the Helmholtz free energy:
\begin{equation}
p(T,V,N)=-\left(\frac{\partial F}{\partial V}\right)_{T,N}.
\end{equation}
When the model is defined on a lattice, the discrete version of this expression should be used. One site of the lattice, which occupies a volume $v_0$, is removed. The change of free energy upon removing this site will be $\Delta F$, and the pressure is
\begin{equation}
p=-\frac{\Delta F}{v_0}.
\label{pd}
\end{equation}
The model defined so far is athermal, since all allowed configurations have the same energy. We consider the chains on the lattice to be composed by monomers, which are localized on lattice sites, connected by polymer bonds placed on lattice edges. The canonical partition function will be given by:
\begin{equation}
Z_n=c_n^{(1)},
\end{equation}
where $c_n^{(1)}$ is the number of walks with $n$ steps, the superscript being used in the literature to denote the constraint to the half-space \cite{b78}. If the site at the wall where the pressure will be calculated, is at a distance $r$ of the grafting point, the restricted partition function, when this site is blocked to the walk, is:
\begin{equation}
Z_n(r)=c_n^{(1)}(r),
\end{equation}
where $c_n^{(1)}(r)$ is the subset of the chains counted in $c_n^{(1)}$ with no monomer on the excluded site. The pressure exerted by the chain on the excluded site may now be calculated using the expression (\ref{pd}) and noticing that $\Delta F=-k_BT[\ln c_n^{(1)}(r)-\ln c_n^{(1)}]$. The result is:
\begin{equation}
p_n(r)=-\frac{k_BT}{v_0}\ln\left(\frac{c_n^{(1)}(r)}{c_n^{(1)}}\right).
\end{equation}

The model may be generalized including an attractive interaction between the chain and the wall. This may be accomplished associating an attractive energy $\epsilon<0$ to each monomer placed on the wall. The unrestricted partition function in this case is:
\begin{equation}
Z_n(\omega)=\sum_{m=0}^n c_n^{(1)}(m)\,\omega^m,
\label{pf}
\end{equation}
where $\omega=\exp(\beta|\epsilon|)$ is the Boltzmann factor associated to each monomer on the wall and $ c_n^{(1)}(m)$ is the number of $n$-step walks with $m$ of the monomers on the wall. The partition function with exclusion of the site is given by:
\begin{equation}
Z_n(\omega,r)=\sum_{m=0}^n c_n^{(1)}(m,r)\,\omega^m,
\label{rpf}
\end{equation}
where now $c_n^{(1)}(m,r)$ is the number of $n$-step walks with $m$ monomers on the wall and with no monomer on the excluded site whose distance to the origin is $r$. The athermal case is recovered for $\omega=1$. In this case, the adsorption transition happens as $\omega$ is increased, between a phase where the monomers are in the bulk of the lattice and another where a non-vanishing fraction of them are on the wall in the thermodynamic limit \cite{dl93}.

One point that might be asked is if the excluded volume effect, which is taken into account for SAWs but neglected in gaussian walks, does have an effect on the asymptotic decay of the pressure with distance. It is well known that, below the upper critical dimension, the critical exponents for polymer models are different from the classical values if the excluded volume effect is taken into account. The results obtained for the model on a semi-infinite square lattice using exact enumerations \cite{j13} seem to provide a negative answer to this question: the asymptotic decay of the pressure with the distance to the grafting point is of the form $1/r^2$, the same obtained for gaussian walks when $d=2$.

The problem was also exactly solved for chains represented by directed walks on a semi-infinite square lattice \cite{rp13}, including the case of interaction with the limiting wall. It was found that below the adsorption transition the pressure decays with $1/r^{3/2}$, but above the adsorption transition it is dominated by a constant term.

In section \ref{model-s} we solve the general model, with interactions between the chain and the wall, on semi-infinite Bethe lattice with a general even coordination number $q \ge 4$. The pressure exerted by the chain on the wall is calculated in section \ref{press}. Final comments and conclusion may be found in section \ref{conc}.

\section{Model and solution on the Bethe lattice}
\label{model-s}
\begin{figure}
\includegraphics[scale=0.6]{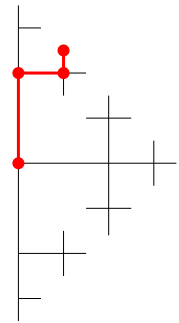}
\caption{Cayley tree with $q=4$ limited by a wall with $q=2$. Four generations of sites are represented. The polymer chain, which starts at the origin, has one monomer on the wall, besides the initial one.} 
\label{model}
\end{figure}

We will study the problem in the canonical formalism. Let us start by defining the semi-infinite Bethe lattice on which the polymeric chain is embedded. The lattice is built starting with a Cayley tree with even coordination number $q$, limited by a wall, which is itself a Cayley tree with $q-2$ first neighbors. The solution of models on the Bethe lattice associated to this Cayley tree may be seen as an approximation to the behavior of these models on hypercubic lattices in $q/2$ dimensions. The lattice is illustrated in figure \ref{model} for the particular case $q=4$. The wall in this case is a $q=2$ Cayley tree, that is, a one dimensional lattice. We notice that the tree shown in this example has four generations of sites, the number of sites in each generation, starting at the central site and moving outwards to sites of a higher generation, is 1, 3, 7, and 19, respectively. We will assume that the polymeric chain is fixed to the central site of the tree. The chain is composed by monomers, represented by circles in the figure, and bonds connecting monomers on first neighbor sites. The monomer placed on the origin is not taken into account, so that the number of monomers is equal to the number of polymeric bonds. Of course this convention has no effect on the results in the thermodynamic limit.
 
The simplest version of the model, studied on the square lattice in \cite{j13} is athermal, but here we will consider an attractive interaction between the chain and the wall, as discussed in the introduction. We notice that, in opposition to what happens on a regular lattice of for directed walks, on the Bethe lattice once the chain leaves the wall it can not get back to it. The basic combinatorial problem we have to solve is to calculate the partition functions of the system (equations (\ref{pf}) and (\ref{rpf})) on the semi-infinite lattice, and thus we need to find the numbers of $n$-step walks with $m$ monomers on the wall $c_n^{(1)}(m)$, as well as the same number when the site at chemical distance $r_c$ of the grafting point $c_n^{(1)}(m,r_c)$. We recall that the chemical distance between two sites on a treelike lattice is the number of steps on the lattice to go from one site to the other, in other words, it corresponds to the difference between the generation numbers of the sites. As will be discussed below, one should distinguish this distance from the euclidean distance between the sites. To avoid surface effects, so that we actually obtain the Bethe lattice solution of the problem, we suppose that the chain does not reach the sites in the last generation of sites (surface) of the tree, so that its size may grow without limit. We expect that the free energy associated to the partition function Eq. (\ref{pf}), in the thermodynamic limit $n \to \infty$ to display a transition between a bulk and an adsorbed phase.

The unrestricted number of configurations $c_n^{(1)}(m)$ may be obtained as follows. For $m=0$, the chain leaves the wall at the first step, so that there is a single possibility for this step and $q-1$ for all the remaining steps, so that:
\begin{equation}
c_n^{(1)}(0)=(q-1)^{n-1}.
\end{equation}
For $m>0$, there are $q-2$ possibilities for the first step, and $q-3$ for the remaining $m-1$ steps on the wall. For the next step, the one where the chain leaves the wall, there will be a single possibility, and for the remaining $n-m-1$ steps there are $q-1$ possibilities, thus:
\begin{equation}
c_n^{(1)}(m)=(q-2)(q-3)^{m-1}(q-1)^{n-m-1}.
\end{equation}
The sum in Eq. (\ref{pf}) may be easily performed, leading to the partition function:
\begin{equation}
Z_n=(q-1)^{n-1}\left\{1+\frac{q-2}{q-3}\left[\frac{x^n-x}{x-1}+
(q-1)x^n\right]\right\},
\label{zn1}
\end{equation}
where $x=\omega(q-3)/(q-1)$.

The contributions to the partition function which are discarded when the site at the wall is excluded may be easily found if we realize that all the forbidden chain configurations have the same first $r_c$ steps, which are on the wall. Therefore, we conclude that:
\begin{eqnarray}
\Delta Z_n(r_c)=Z_n-Z_n(r_c)=
\nonumber \\
\left(\frac{x}{q-3}\right)^{r_c} (q-1)^ {n-1}
\left[\sum_{m=0}^{n-r_c-1}x^m+(q-1)x^{n-r_c}\right],
\end{eqnarray}
where we assume $n \ge r_c+2$. This restriction is not important because basically we will be interested in the thermodynamic limit $n \to \infty$, with a fixed value of $r_c$. Performing the sum, the partition function for the model with the excluded site will be:
\begin{eqnarray}
Z_n(r_c)&=&(q-1)^{n-1}\left\{1+\frac{q-2}{q-3}\left[\frac{x^n-x}{x-1}+
(q-1)x^n\right] \right.
\nonumber \\
&&\left.-\left(\frac{1}{q-3}\right)^{r_c}\left[\frac{x^{n}-x^{r_c}}{x-1}+(q-1)x^n
\right]\right\}.
\label{znr1}
\end{eqnarray}

We may now obtain the thermodynamic properties of the model, we will start with the case where no site is excluded. Using the partition function Eq. (\ref{zn1}), the dimensionless Helmholtz free energy per monomer of the chain will be:
\begin{equation}
\phi=\frac{f}{k_BT}=-\lim_{n \to \infty}\frac{1}{n}\ln(Z_n).
\end{equation}
The partition function is a product of two factors. In the thermodynamic limit, one of the factors will  dominate, depending of the magnitude of $x$. At low values of $x$ the polymer is located in the bulk. At $x=1$ a discontinuous phase transition happens and the polymer is completely adsorbed on the wall. This is shown by the result:
\begin{equation}
\phi=
\left\{ \begin{array}{ll}
-\ln(q-1) & \mbox{for $x \le  1$,} \\
-\frac{|\epsilon|}{k_BT}-\ln(q-3) & \mbox{for $x>1$.}
\end{array}
\right.
\end{equation}
In other words, the system may be in two phases: a bulk phase at high temperatures and an adsorbed phase at low temperatures. The temperature where both phases coexist is:
\begin{equation}
T_c=\frac{|\epsilon|}{k_B\ln\frac{q-1}{q-3}}.
\end{equation}
Other thermodynamic functions may be readily obtained. The entropy per monomer is:
\begin{equation}
\frac{s}{k_B}=
\left\{ \begin{array}{ll}
\ln(q-1) & \mbox{for $T \ge  T_c$,} \\
\ln(q-3) & \mbox{for $T < T_c$;}
\end{array}
\right.
\end{equation}
and the internal energy vanishes above $T_c$ and is equal to $\epsilon$ below the transition temperature. In opposition to what is found on regular lattices \cite{dl93} and for other restricted self-avoiding walks such as directed walks \cite{p88}, where the adsorption transition is continuous, on the Bethe lattice with a wall as defined here, it is a totally discontinuous transition, between a phase where the fraction of monomers on the wall vanishes and another where it is equal to unity. This is due to the fact that, in the present model, when the chain leaves the wall it is not allowed to return to it.

\section{Pressure exerted on the wall}
\label{press}
The pressure applied by the polymer on the wall may now be calculated using the Eq. (\ref{pd}) discussed in the introduction. As a site on the wall located at a chemical distance $r_c$ from the origin is removed, the change in free energy will be:
\begin{equation}
\Delta F=-k_B \ln \frac{Z_n(r_c)}{Z_n}.
\end{equation}
It is worth remarking that the density of monomers at the excluded site is given by $\rho(r_c)=1-Z_n(r_c)/Z_n$ so that we have the state equation:
\begin{equation}
p_n=-\frac{k_BT}{a^{q/2}}\ln[1-\rho(r_c)]
\end{equation}
between the pressure and the local density of monomers. We have replaced the volume of the elementary cell of the lattice by $a^{q/2}$, where $a$ is the lattice parameter and $q/2$ is the number of orthogonal directions which meet at each lattice site, as will be discussed in more detail below. Using the Eqs. (\ref{zn1}) and (\ref{znr1}) for the partition function without and with site exclusion, respectively, the dimensionless pressure $\pi_n(r_c)=p_n(r_c)a^{q/2}/(k_BT)$ applied by a chain with $n$ monomers is:
\begin{equation}
\pi_n(r_c)=-\ln\left\{1-\frac{\left(\frac{1}{q-3}\right)^{r_c}\left[\frac{x^{n}-x^{r_c}}{x-1}+(q-1)x^n\right]}{1+\frac{q-2}{q-3}\left[\frac{x^n-x}{x-1}+(q-1)x^n\right]}\right\},
\label{pin}
\end{equation}
where we have replaced 
The pressure in the thermodynamic limit $\pi=\lim_{n \to \infty} \pi_n$ may now be obtained. The result is:
\begin{equation}
\pi=
\left\{ \begin{array}{ll}
-\ln\left[ 1-\frac{\left(\frac{x}{q-3}\right)^{r_c}}{1+\frac{x}{q-3}}\right]&
\mbox{for $T \ge  T_c$,} \\
-\ln\left[ 1-\frac{(q-3)^{1-r_c}}{q-2}\right] 
& \mbox{for $T < T_c$.}
\end{array}
\right.
\label{pi}
\end{equation}
As expected for a field-like thermodynamic quantity, the pressure is continuous at the transition. In the adsorbed phase the dimensionless pressure does not change with temperature, in other words, the pressure is proportional to the temperature. This is consistent with the fact that the adsorption of the chain on the wall is complete at any temperature below the transition.

\begin{figure}
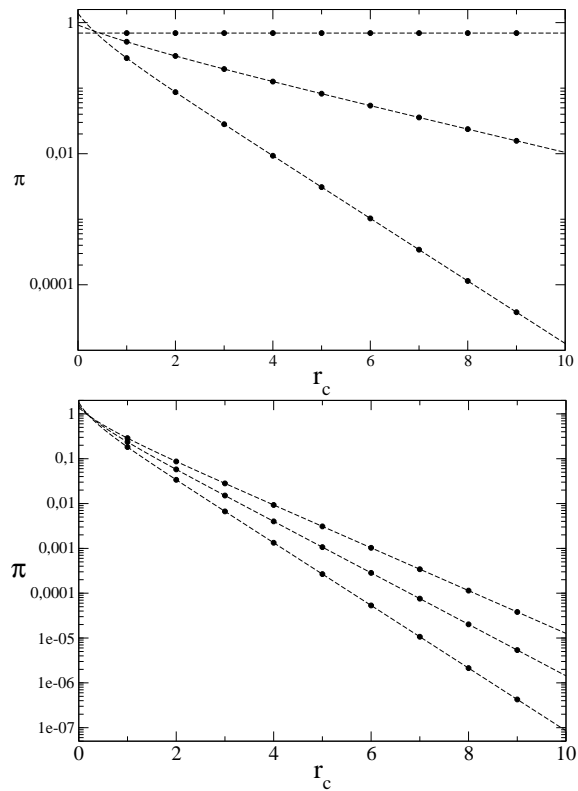

\includegraphics[scale=0.3]{pq4.eps}
\includegraphics[scale=0.3]{pq6.eps}
\caption{Pressure as a function of chemical distance. The dashed lines are guides to the eye. Pressure increases as $\omega$ is increased, up to the coexistence value $\omega_c$. Upper graph: $q=4$, $\omega=1,\;2$ and $\omega=\omega_c=3$. Lower graph: $q=6$, $\omega=1,\;4/3$ and $\omega=\omega_c=5/3$.}
\label{pressg}
\end{figure}

The asymptotic behavior of the pressure for $r_c \gg 1$ may be found from the general expressions in Eq. (\ref{pi}). Above the transition temperature, the result is:
\begin{equation}
\pi \approx 
\frac{\left(\frac{x}{q-3}\right)^{r_c}}{1+\frac{x}{q-3}},
\label{pia}
\end{equation}
while below the transition temperature, for $q=4$ the pressure is constant and for larger values of $q$ we have
\begin{equation}
\pi \approx \frac{(q-3)^{1-r_c}}{q-2}.
\label{pib}
\end{equation}
It is worth noting that for directed walks on the square lattice the pressure does also not decay with the distance at and above the transition temperature, but it is temperature dependent \cite{rp13}
\begin{equation}
\pi = \log(\omega-1).
\end{equation}
In figure \ref{pressg} we present the pressure as a function of the chemical distance for $q=4$ and $q=6$ and some values of the Boltzmann weight $\omega$, up to the coexistence value $\omega_c$, since for $\omega>\omega_c$ the pressures are identical to the ones at coexistence. The qualitative difference between results for $q=4$ and for $q>4$ is apparent: in the first case the pressure does not change with distance for $\omega \ge \omega_c$, while it does decay exponentially for larger coordination numbers. We see that the decay of the pressure with the chemical distance is exponential, in opposition with the power law decay found for SAWs on the square lattice \cite{j13} and for directed walks \cite{rp13}.

\begin{figure}
\includegraphics[scale=0.3]{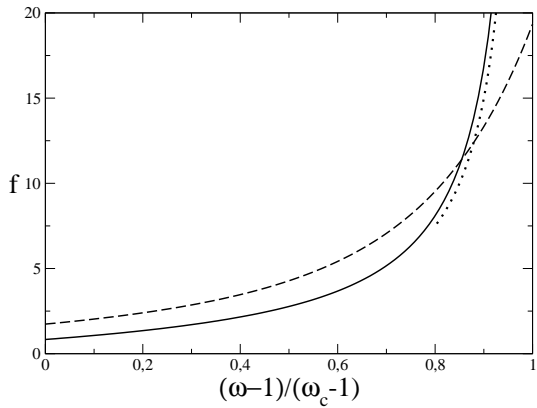}
\caption{Force exerted by the chain on the wall as a function of $(\omega-1)/(\omega_c-1)$ for $q=4$ (full line) and $q=6$ (dashed line). The dotted line corresponds to the asymptotic behavior of the force for $q=4$ close to the adsorption transition.}
\label{force}
\end{figure}

Another quantity of interest is the total force applied by the walk on the wall. This quantity may be obtained summing the pressures for all sites of the wall other than the central site. The force will be:
\begin{equation}
{\cal F}_n(x)=\frac{k_BT}{a}(q-2)\sum_{r_c=1}^n(q-3)^{r_c-1}\pi_n(x),
\end{equation}
where $\pi_n(x)$ is given by Eq. (\ref{pin}). Although we were not able to sum the pressures analytically, it is not difficult to find a precise numerical value for the force in the thermodynamic limit ${\cal F}=\lim_{n \to \infty}{\cal F}_n$. Some results are presented in figure \ref{force}, where the dimensionless force $f={\cal F}a/(k_BT)$ is shown as a function of $(\omega-1)/(\omega_c-1)$ for $q=4$ and $q=6$. For $q=4$ the force in the athermal case $\omega=1$ is $f \approx 0.8336034$ and as the adsorption transition is approached, it diverges, since the pressure does not decay with the distance in this limit. Using the asymptotic expression Eq. (\ref{pia}) for the pressure, we may find that, close to the adsorption transition $\omega \to 3_-$, the force behaves as:
\begin{equation}
f \approx \frac{1}{1-\omega/3}.
\end{equation}
The dotted curve in figure \ref{force} corresponds to this limiting behavior.
For the lattice with $q=6$, the force at $\omega=1$ is $f=1.736920$ and its maximum for $\omega \ge \Omega_c=5/3$ assumes the value $f=19.34622$. It is interesting to compare the value for $q=4$ in the athermal limit with the estimate obtained for SAWs on the square lattice, which is $f_{SAW} \approx 1.533$, larger than the Bethe lattice result. Integrating the pressure in Eq. (\ref{gauss}) for gaussian walks, the result is $f_G=1$ \cite{j13}.

\section{Final discussions and conclusion}
\label{conc}
It should be noticed that it is misleading to associate the chemical distance between two sites on the Bethe lattice to the euclidean distance between these sites. This has lead to some confusion in the literature, such as apparent non-classical critical exponents for SAWs on this lattice \cite{hu98}. One consistent way to define the euclidean distance between two sites is to embed the Bethe lattice in a hypercubic lattice, so that at each new generation of sites new bonds in directions which are orthogonal to all previous directions are added \cite{sca00}. This lattice will be infinite dimensional in the thermodynamic limit, consistent with the geometric definition of the dimensionality of this lattice  and with the fact that the exact solution of statistical mechanical models which display continuous phase transitions on this lattice lead to classical critical exponents,  similar to what happens with the Curie lattice, where each  site interacts with all others and the exact solution of models correspond to mean-field approximations of the same models on regular lattices, the exact solutions on the Bethe lattice is equal to the Bethe approximation of the same models on regular lattices with the same coordination number \cite{b82}.

If the euclidean distances on the Bethe lattice are defined as suggested in \cite{sca00}, the euclidean distances between sites with the same chemical distance may be different, since the euclidean distance will depend of the number and localization of the bends in the walk between the sites, so that the pressure at sites with the same euclidean distance may be different. For example, two sites connected by a walk with three steps in a direction and four steps in a orthogonal direction ($r_c=7$) will be  at the same chemical distance as two sites linked by a $5$-step walk in the same direction. Therefore, there is no simple way to associate a single pressure to a chemical distance. In general, if the chemical distance between two sites is $r_c$, the euclidean distance between them is in the range $[\sqrt{r_c},r_c]$, and the average of the square of the euclidean distances $r$ between these sites, on a Bethe lattice with coordination number $q$,  is \cite{sca00}:
\begin{equation}
\langle r² \rangle=\frac{2(q-1)}{(q-2)^2}\left[(q-2)r_c-1+\frac{1}{(q-1)^{r_c}}\right]-r_c,
\end{equation}
where it is apparent that for large values of $r_c$ we have $\langle r^2 \rangle \approx r_c$, so that the mean-field (random walk) value $\nu=1/2$ is recovered for the critical exponent. Of course all this discussion is valid only for the case $q>4$. If the semi-infinite lattice has coordination number $q=4$, the wall will have $q^\prime=2$ and will be a one-dimensional lattice, for which $r_c=r$.

\begin{acknowledgments}
We acknowledge Tiago J. de Oliveira for a critical reading of the manuscript.
\end{acknowledgments}

\end{document}